\documentclass[aps,preprint,preprintnumbers,amsmath,amssymb,amsfonts,groupedaddress,superscriptaddress,showpacs,showkeys,floatfix]{revtex4-1}

\hfuzz=\maxdimen
\usepackage[dvipsnames]{xcolor}
\usepackage{graphicx}
\usepackage{subfigure}
\usepackage{mathtools}
\usepackage{bm}
\usepackage{color}
\usepackage{array}
\usepackage{cases}
\usepackage{upgreek}

\newcommand{\bmath}{\begin{mathletters}}
\newcommand{\emath}{\end{mathletters}}
\newcommand{\be}{\begin{eqnarray}}
\newcommand{\ee}{\end{eqnarray}}
\newcommand{\ba}{\begin{array}}
\newcommand{\ea}{\end{array}}

\newcommand{\no}{\nonumber}

\newcommand{\De}{\Delta}

\newcommand{\w}{\omega}

\newcommand{\ga}{\gamma}
\newcommand{\Ga}{\Gamma}

\newcommand{\pr}{\prime}

\newcommand{\dprime}{\prime\prime}
\newcommand{\rtar}{\rightarrow}

\newcommand{\intt}{\int_{0}^{t}}

\newcommand{\calI} {\mathcal I}

\newcommand{\calL} {\mathcal L}
\newcommand{\calP} {\mathcal P}

\newcommand{\calS} {\mathcal S}
\newcommand{\calT} {\mathcal T}
\newcommand{\calU} {\mathcal U}

\newcommand{\calW} {\mathcal W}

\newcommand{\rmB} {\mathrm{B}}

\newcommand{\rmI} {{\mathrm{I}}}

\newcommand{\rmR} {{\mathrm{R}}}
\newcommand{\rmS} {\mathrm{S}}

\newcommand{\rmX} {\mathrm{X}}

\newcommand{\Tr} {\mathrm{Tr}}

\newcommand{\RDM} {\mathrm{RDM}}
\newcommand{\tot} {\mathrm{tot}}
\newcommand{\eq}  {\mathrm{eq}}

\newcommand{\CI} {\mathrm{CI}}
\newcommand{\fit} {\mathrm{fit}}

\begin{document}
\title{Linear Absorption Spectrum of the Spin-Boson Model Studied by Extended Hierarchical Equations of Motion}
\author{Qianlong Wang}
\affiliation{Department of Physics,  Zhejiang University, Hangzhou, Zhejiang, 310027, China}

\author{Jianlan Wu}
\email{jianlanwu@zju.edu.cn}
\affiliation{Department of Physics,  Zhejiang University, Hangzhou, Zhejiang, 310027, China}

\begin{abstract}

With a decomposition scheme for the bath correlation function, the
hierarchical equations of motion (HEOM) are extended to the zero-temperature
sub-Ohmic and Ohmic spin-boson model.  We investigate the linear absorption
spectrum of the sub-Ohmic and Ohmic spin-boson model at zero temperature.  By
applying the extended HEOM, the equilibrium spin dynamics are obtained
approximately.  Then the linear response function is calculated according to
the Kubo formula rewritten in Liouville space.  To explore the essence of
phase transition, we compute the linear absorption spectrum defined by the
linear response function of a dipole moment.  By analyzing the peak
structure of the linear response spectrum, we get the dependence of linear
absorption spectrum with Kondo parameter and different bath exponents.  The
spin relaxation dynamics are also calculated to explore the
coherent-incoherent dynamic transition (CI) and delocalized-localized phase
transition (DL).  The corresponding phase diagram of DL and CI transition are
also obtained.  We pose a energy level picture to understand the different
mechanism of DL phase transition between the deep sub-Ohmic and Ohmic
spin-boson model.
\end{abstract}

\maketitle

\section{Introduction}
\label{sec1}
As one of the most generally studied models in open quantum system, the spin-boson model (SBM) has
attracted significantly attention because it can capture the essential physics
of quantum decoherence and quantum phase transition~\cite{Leggett1987:RMP}.
Different environment can induce abundant physical phenomenons.
Optical spectroscopy is a powerful tool to understand the abundant phenomenons
behind their electronic structure and dynamics.
It has been widely used to investigate the excitonic dynamics, such as
biological light harvesting systems~\cite{renger2001:PR,grondelle2006:PCCP,cheng2009:PRC,Herman2018:JPB}.

Physically, the linear absorption spectrum can describe the single-photon
absorption process, which determines how the energy level is affected by the weak external perturbation.
The investigation of the linear absorption spectrum relies on the equilibrium state of total system.
For the open quantum system interacting with an environment bath, it is still
challenging to calculate the equilibrium state, mostly due to the huge number of
the bath degrees of freedom.
In order to deal with this challenge, numerous methods has been developed~\cite{Berkelbach2017:JCP}.
One common strategy is to apply the rigorous variational principle.
For example, the variational polaron transformation and its extensions
~\cite{Silbey1984:JCP,Zheng2007:PRB,Zheng2009:PRB,Zheng2009:PRE, Zheng2013:JCP,Plenio2011:PRL,
Zhao2016:JCP,Chen2018:PRB}
that use a self-consistent way to determine a renormalized tunneling amplitude and the equilibrium state,
the (multilayer) multiconfiguration time-dependent Hartree (MCTDH/ML-MCTDH)
method~\cite{Cederbaum1990:CPL,Wang2001:hybridP1,Wang2001:hybridP2,Wang2003:MLMCTDH,Wang2008:NJP}
obtained via the Dirac-Frenkel variational principle,
and the time-dependent variational principle (TDVP) for variational matrix product states (VMPS)~\cite{Chin2016:PRB}.
Meanwhile, many other kinetic methods based on the evolution of wave function or
density matrix can also be used to calculate the equilibrium properties.
For example, the Hilbert space can be efficient compressed in the methods of
the numerical renormalization group (NRG)~\cite{Schiller2005:PRL, Vojta2007:PRL, Freyn2009:PRL, Costi1996:PRL, Anders2007:PRL, Bulla2007:PRB},
the sparse polynomial space representation (SPSR)~\cite{Fehske2009:PRL},
the time-dependent density matrix renormalization group (t-DMRG)~\cite{Plenio2010:PRL} , etc.
Additionally,
the bosonic Hilbert space can be alternatively sampled by stochastic trajectories in the methods of
the quantum Monte Carlo (QMC)~\cite{Mak1994:PRB,Bulla2009:PRL},
the path-integral Monte Carlo (PIMC)~\cite{Escher2004:JCP,Ankerhold2013:PRL,Muelken2013:arXiv},
the stochastic Liouville-von Neumann equation (SLN)~\cite{Stockburger2004:CP},
the stochastic path integral (SPI)~\cite{Moix2012:JCP, Moix2012:PRB}, etc.
Due to the limitation of space, a huge number of other methods are not able to be discussed here~\cite{Berkelbach2017:JCP}.

In this paper, we applied the HEOM formalism to calculate the linear absorption spectrum based on the
linear response function.
As a numerical exact kinetic method, HEOM has been widely used in open quantum physics~\cite{Yan2009:JCP,Yan2011:arXiv,Shi2013:JCP,Tanimura2020:JCP}.
With the assistance of the auxiliary density operator, all the bath effects exert
on the reduced system dynamics are contained.
The HEOM formalism was firstly proposed by Tanimura and Kubo in the case of Drude
spectral density and high temperature~\cite{Tanimura1989:JPSJ}.
In order to overcome the limitation of original HEOM in exponential, many
modifications have been proposed to improve the application width of HEOM.
To improve the calculation efficiently of high truncation order case, Q. Shi \textit{et al.}
rescaled the HEOM equation and proposed a filtering algorithm which can significantly
reduce the number of auxiliary density operators ~\cite{Shi2009:JCP}.
To deal with any complex form of the bath time correlation function, we proposed
an extended form of HEOM. By decomposing the bath time correlation function into
a series of basis function, we can deal with arbitrary bath time correlation function
in principle~\cite{Wu2015:eHEOM,Duan2017:PRB,Duan2017:JCP,Wang2019:JCP}.
In addition, recently, some new bath correlation decomposition schemes have been proposed to improve
the original exponential HEOM scheme, such as the Chebyshev
hierarchical equations of motion (C-HEOM)~\cite{Rahman2019:JCP}.
With the assistance of HEOM method,
Tanimura and coworkers~\cite{Kato2004:JCP,Tanimura2006:JPSJ,Ishizaki2006:JPSJ,Ishizaki2007:JPC}
have applied it to calculate the two-dimensional infrared spectra.

The rest of this paper is organized as follows.
In Sec.~\ref{sec2}, the Hamiltonian of the spin-boson model is introduced, followed by the description
of the bath spectral density and time correlation function. In a finite time
interval, the bath time correlation is fitted well with a series of base functions.
The Kubo formula in Liouville space is introduced to
calculate the linear response function and the linear absorption spectrum.
In Sec.~\ref{sec3}, we introduce how to use the extended HEOM to calculate the
equilibrium state and evolution of the perturbed auxiliary density operators.
As for the extended HEOM, the technique of decomposition is applied to the bath correlated functions with
respect to the Ohmic and sub-Ohmic spectral density.
In Sec.~\ref{sec4}, we calculate the spin dynamics and linear response function with
the extended HEOM at zero temperature both for sub-Ohmic and Ohmic SBM.
Through an analysis of the average magnetic moment and its rate kernel,
we obtain the DL phase transition and the CI dynamic transition.
Meanwhile, different behaviors of the linear absorption spectrum between the deep sub-Ohmic and
Ohmic SBM are presented.
A summary of this work is given in Sec.~\ref{sec5}.

\section{Linear Absorption Spectrum of the Spin-Boson Model}
\label{sec2}

\subsection{Hamiltonian}
\label{sec2a}

The Hamiltonian of the spin-boson model is written as \cite{Leggett1987:RMP}
\be
  H &=& H_\rmS+H_\rmB+H_\mathrm{SB} \no \\
  &=& \Delta\sigma_{x} + \frac{1}{2}\sum_j\left(
  p_{j}^{2} + \omega_{j}^{2}q_{j}^{2}
  \right) +\sigma_{z}\sum_{j}c_{j}q_{j}.
  \label{eq:01}
\ee
The system Hamiltonian, $H_\rmS = \De \sigma_x$, describes the tunneling between two degenerate spin states
($|+\rangle$ and $|-\rangle$), where $\Delta$ is the tunneling amplitude and $\{\sigma_x, \sigma_y,
\sigma_z\}$ is the set of the Pauli matrices.  The bath Hamiltonian, $H_\rmB = (1/2)\sum_j(p_{j}^{2} +
\omega_{j}^{2}q_{j}^{2})$, is introduced for an ensemble of harmonic oscillators, where $q_j$, $p_j$ and
$\omega_j$ are the coordinate, momentum and frequency of the $j$-th oscillator. The system-bath interaction,
$H_\mathrm{SB}= \sigma_zF_\mathrm{B}=\sigma_{z}\sum_{j}c_{j}q_{j}$, follows a bilinear form, where $c_j$ is
the amplitude between the spin and the $j$-th oscillator. The mass of each oscillator is set to be unity
($m_j=1$) and the same for the reduced Planck constant ($\hbar\!=\!1$).

\begin{figure}[tp]
\includegraphics[width=0.65\columnwidth]{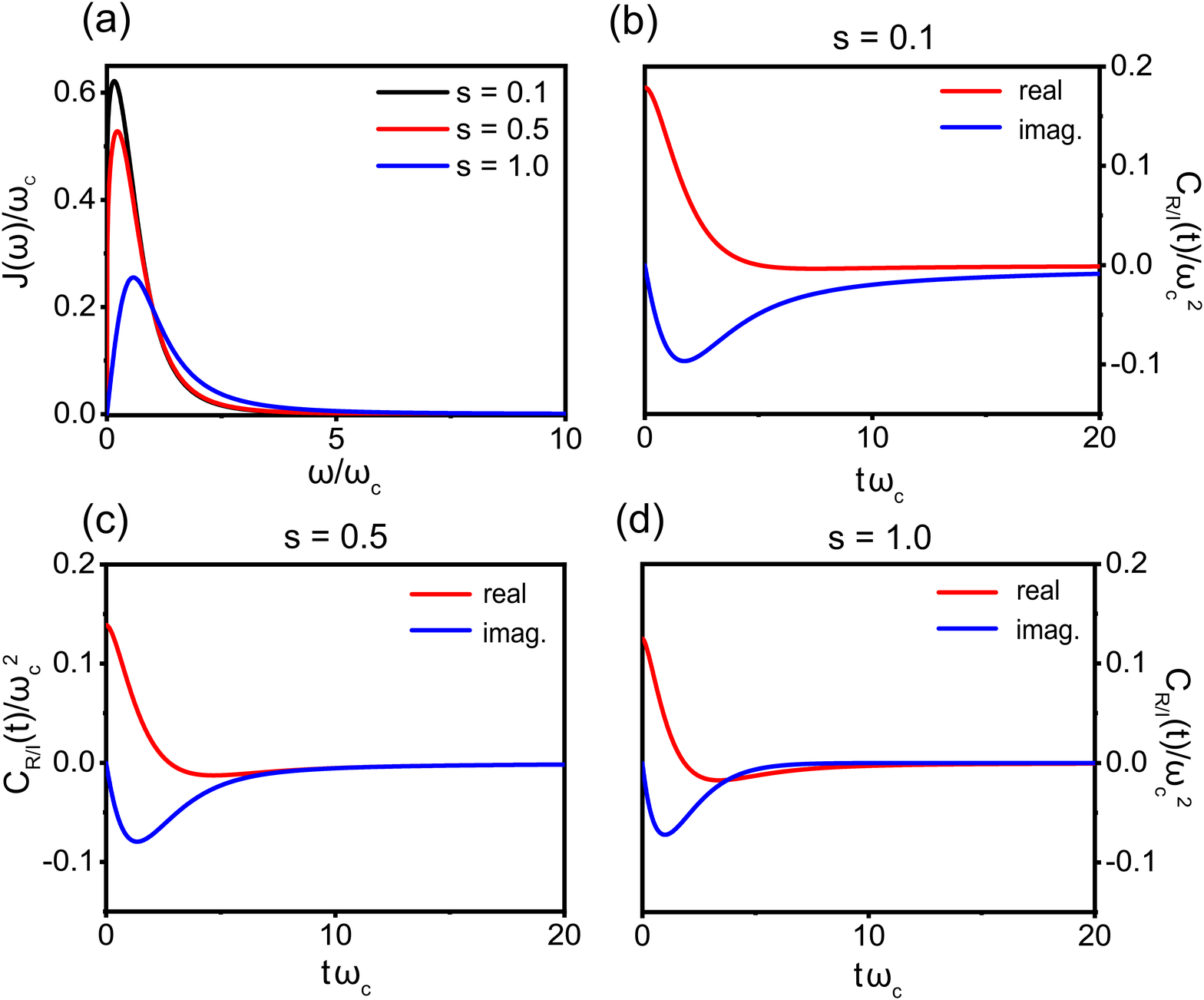}
\caption{(a) The spectral densities $J(\omega)$ of the sub-Ohmic and Ohmic baths for three different exponents,
$s \!=\! 0.1$, $0.5$ and $1$. The cutoff function follows a rational fraction form.
(b)-(d) The corresponding bath correlation functions $C(t)$ for the three values of $s$ at $T=0$.
The red and blue lines denote the real and imaginary parts of the $C(t)$,
respectively. The Kondo parameter is set to be $\alpha \!=\! 0.5$.
}
\label{fig:01}
\end{figure}

The influence of the bosonic bath onto the spin system is characterized by a spectral density,
$J(\omega) = (\pi/2)\sum_j(c_j^2/\omega_j)\delta(\omega-\omega_j)$. Based on its low-frequency feature,
the spectral density can be expressed as
\be
J(\omega) = \frac{\pi}{2}\alpha\omega^{s}\omega_{c}^{1-s}f\left(\frac{\omega}{\omega_{c}}\right),
\label{eq:02}
\ee
where the Kondo parameter $\alpha$ represents the average system-bath coupling strength.
The bosonic bath is categorized by the exponent $s$: Ohmic ($s=1$), sub-Ohmic ($0<s<1$)
and super-Ohmic ($s>1$)~\cite{Leggett1987:RMP}. The cutoff function $f(\omega/\omega_{c})$
is assigned for the high frequency feature of the spectral density where $\omega_c$ is
a cutoff frequency. In this paper, we focus on a rational fraction form~\cite{Shao2008:JCP,Wang2019:JCP},
\be
f\left(x=\frac{\omega}{\omega_{c}}\right) = \frac{1}{(1+x^{2})^{2}},
\label{eq:03}
\ee
while the other forms of $f(x)$ can be studied similarly~\cite{Duan2017:JCP,Wang2019:JCP}.

With respect to a Gaussian distribution $\rho_\rmB \!\propto\! \exp(-\beta H_\rmB)$,
the influence of the bosonic bath is alternatively described by the time correlation function,
$C(t) = \Tr_\rmB\{F_\rmB(t)F_\rmB\rho_\rmB\}$. 
The parameter $\beta\!=\!1/{k_B T}$ is the inverse product of the Boltzmann constant $k_B$ and temperature $T$.
With the substitution of the spectral density, the bath correlation function is explicitly given by
\be
C(t) &=& C_\rmR(t)+i C_\rmI(t) \no \\
&=& \frac{1}{\pi} \int_{0}^{\infty}d\omega J(\w)\left[ \coth \frac{\beta\w}{2}\cos \w t - i \sin \w t\right],
\label{eq:04}
\ee
where $C_{\rmR}(t)$ and $C_{\rmI}(t)$ are the real and imaginary parts, respectively.
In this paper, we focus on zero temperature ($T\!=\!0$) where the Bose factor is  $\coth(\beta\omega/2)\!=\!1$.
In Fig.~\ref{fig:01}(a), we present the curves of $J(\omega)$ with  $s\!=\!0.1,0.5$ and $1$ as a demonstration.
The corresponding time correlation functions are shown in Figs.~\ref{fig:01}(b)-\ref{fig:01}(d).
For each exponent, the real part $C_{\rmR}(t)$ experiences the change from a positive to negative value
with the increase of the time while the imaginary part $C_\rmI(t)$ is always negative. For the sub-Ohmic
case (e.g., $s=0.1$ and 0.5), both $C_\rmR(t)$ and $C_\rmI(t)$ behaves asymptotically as
$C_{\rmR/\rmI}(t\rightarrow \infty)\sim -(\omega_ct)^{-(1+s)}$ in the long time limit.
For the Ohmic case ($s=1$), the asymptotic behavior of the real part is the same,
$C_{\rmR}(t\rightarrow \infty)\sim -(\omega_ct)^{-2}$ while the imaginary part is
analytically given by $C_\rmI(t)=-(1/8)\pi \alpha \omega_c^3 t e^{-\omega_c t}$.

\subsection{Linear Response Function and Linear Absorption Spectrum}
\label{sec2b}

The physical properties of the spin-boson model are determined by the eigenstructure
of the Hamiltonian $H$ in Eq.~(\ref{eq:01}).
One approach of calculating the eigenstates is to diagonalize $H$
by either a deterministic solver or random sampling. Alternatively, we can introduce
an external field and inspect the excited states through the linear response of
the spin system.

With respect to a time-dependent electric field $\vec{E}(t)$,
the total Hamiltonian is approximated as
\be
H_{\tot}(t) = H  - \sigma_{z}\mu_E E(t),
\label{eq:05}
\ee
where $\mu_E$ is the component of the dipole moment $\vec{\mu}$ along the direction of $\vec{E}(t)$.
For convenience, $\mu_E$ is set to be the unity ($\mu_E=1$) throughout this paper.
The linear response function $\chi(t)$ is then derived as~\cite{Mukamel1995}
\be
\chi\left(t\right) & =  i\Uptheta (t) \Tr \{\left[ \sigma_z(t), \sigma_z \right]\rho_{\eq}\},
\label{eq:06}
\ee
where $\Uptheta (t)$ is the Heaviside step function, $\Tr \!=\! \Tr_{\rmS}\Tr_{\rmB}$
denotes a trace over the degrees of freedom of both the system and the bath,
$\sigma_z(t) \!=\! \exp(iH t)\sigma_z\exp(-iH  t)$ is the time-dependent Pauli-$Z$
matrix, and $\rho_{\eq} \!\propto\! \exp(-\beta H)$ is the equilibrium density
matrix of the spin-boson model without the influence of $\vec{E}(t)$.
To be consistent with our HEOM calculation, Eq.~(\ref{eq:06}) is re-formulated
in the Liouville space, given by
\be
\chi\left(t\right)  &=&  i \Uptheta(t)\Tr\{\sigma_{z} \exp(-i\calL t)\calL_z \rho_\eq\} \no \\
&=& i \Uptheta(t)\Tr_\rmS\{\sigma_{z} \varrho_\rmS(t)\}.
\label{eq:07}
\ee
For a Liouville superoperator $\calL=[H, \cdots]$, its time evolution
superoperator is given by $\exp(-i\calL t)=\exp(-i H t)\cdots \exp(i Ht)$.
Based on an initial value $\varrho(0)=\calL_z\rho_\eq=[\sigma_z, \rho_\eq]$, we introduce a
quasi reduced density matrix (RDM) as
\be
\varrho_\rmS(t)=\Tr_\rmB\{\exp(-i\calL t)\varrho(0)\}.
\label{eq:07A}
\ee
As a result, one procedure of calculating the linear response function is described as
\be
\rho(0) &&\overset{\lim_{t^\pr\rightarrow\infty}\exp\left(-i\calL t^{\prime}\right)\rho(0)}{\xrightarrow{\hspace*{3cm}}} \rho_{\eq}
\overset{\calL_{z}\rho_{\eq}}{\xrightarrow{\hspace*{1cm}}} \varrho(0) \no \\
&&~\overset{\Tr_\rmB\{\exp(-i\calL t)\varrho(0)\}}{\xrightarrow{\hspace*{3cm}}} \varrho_{\rmS}(t)
\overset{\Tr_{\rmS}\{\sigma_{z}\varrho_{\rmS}(t)\}}{\xrightarrow{\hspace*{2cm}}}\chi(t). \no
\ee
(1) The spin-boson model is propagated from an arbitrary initial state $\rho(0)$ (e.g., a product state)
to the equilibrium state $\rho_\eq=\lim_{t^\pr\rightarrow\infty}\exp(-i\calL t^\pr)\rho(0)$.
(2) The equilibrium state is taken the action of the Pauli-$Z$ matrix to form
a non-normalized state $\varrho(0)=\calL_z\rho_\eq$.
(3) The quasi RDM $\varrho_\rmS(t)$ is obtained through the time evolution in Eq.~(\ref{eq:07A}).
(4) The linear response function $\chi(t)$ is estimated by Eq.~(\ref{eq:07}).
The real-time HEOM will be utilized in steps (1) and (3).
Furthermore, we further interested in the linear absorption spectrum,
$\chi^{\dprime}(\omega)$, which can be viewed as an experimental measurement.
Based on its definition, the linear absorption spectrum is given by
\be
\chi^{\dprime}(\omega)=\operatorname{Im} \int_{-\infty}^{\infty}\chi(t) e^{i\omega t}\,dt,
\label{eq:08}
\ee
where $\mathrm{Re}$ and $\mathrm{Im}$ denote the real and imaginary parts of a complex
variable, respectively.

\section{An Extended HEOM}
\label{sec3}

In this section, we provide a brief description of an extended
HEOM~\cite{Wu2015:eHEOM,Duan2017:PRB,Duan2017:JCP} for the calculation of the
linear response function and the linear absorption spectrum. Based on the
decomposition of the bath correlation function $C(t)$, the HEOM builds a
linearized array of time differential equations of the auxiliary density
operators (ADOs), which gives rise to the RDM at the zeroth order. Our
decomposition is performed onto $C(t)$ directly rather than the Bose factor and
the spectral density so that the HEOM can be easily extended to the zero
temperature ($T=0$). More importantly, the HEOM is founded on a complete dynamic
set of ADOs after tracing the bath degrees of freedom so that this method can be
applied to the numerical calculation of dynamic quantities beyond the RDM. Many
efforts have been attributed to the development of systematic and exact
theoretical frameworks~\cite{Tanimura2012:JCP,Tanimura2015:JCP,Wang2013:PRB}.

\begin{figure}[tp]
\includegraphics[width=0.65\columnwidth]{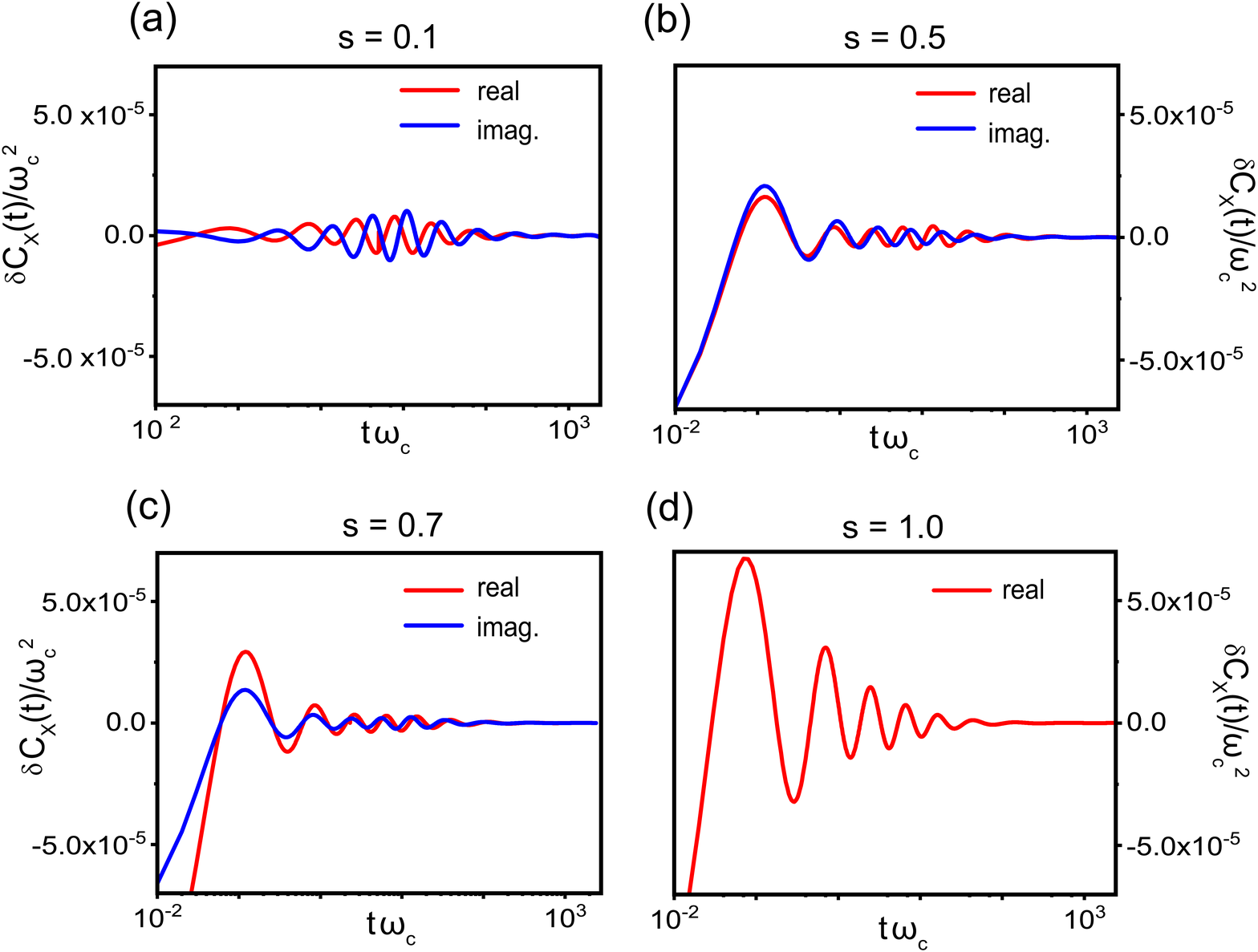}
\caption{The instantaneous errors between the fitting results of $C_{\rmX=\rmR/\rmI}(t)$
and their exact values. The bath exponents in four panels are (a) $s\!=\!0.1$, (b) $s\!=\!0.5$,
(c) $s\!=\!0.7$ and (d) $s\!=\!1$. The line with red and blue denote the results of the real
and imaginary parts, respectively. The Kondo parameter is set to be $\alpha \!=\! 0.5$.}
\label{fig:02}
\end{figure}

\subsection{Decomposition of the Bath Correlation Function}
\label{sec3a}

To treat a general bath at an arbitrary temperature,
an efficient and reliable decomposition of the bath correlation function
is a key technical step in the application of our extended HEOM.
In principle, $C(t)$ can be decomposed exactly over a complete
set of orthonormal basis functions, which is however inefficient in the scenario of
a slowly varying function. On the other hand,
the correlation functions of both the sub-Ohmic and Ohmic baths
exhibit a similar behavior: a non-monotonic variation in the short time regime while an asymptotic power-law
decay in the long time regime. 
Accordingly, we assume the following function,
\be
C^{\fit}(t) =  \sum_{n=1}^{N_\rmR} a_{\rmR;n}\varphi_{\rmR;n}(t) + i \sum_{m=1}^{N_\rmI} a_{\rmI;m}\varphi_{\rmI;m}(t),
\label{eq:10}
\ee
to fit the bath correlation function over a large but finite time interval ($0\leqslant t\leqslant t_{\max}$).
At the maximum time $t_{\max}$, the spin system is assumed to be close to its final equilibrium state.
Here $\{\varphi_{\rmR;n}(t), a_{\rmR;n}\}$ and $\{\varphi_{\rmI;m}(t), a_{\rmI;m}\}$ are two sets of basis functions
and their coefficients for the real and imaginary parts, respectively. In practice, we choose oscillatory and
nonoscillatory exponentially decaying functions~\cite{Duan2017:PRB,Duan2017:JCP,Wang2019:JCP},
\be
\{\varphi_{\rmX;n}(t)\} &=&
\{
\cos(\w_{\rmX; 1} t)e^{-\ga_{\rmX; 1} t},
\sin(\w_{\rmX; 1} t)e^{-\ga_{\rmX; 1} t},
\cdots,
e^{-\Ga_{\rmX;1}t}, e^{-\Ga_{\rmX; 2}t}, \dots
\},
\label{eq:11}
\ee
with $\rmX\!=\!\rmR$ and $\rmI$. The number of the basis functions, $N_\rmR$ and $N_\rmI$,
are allowed to be different.
In the case of $C_\rmI(t; s=1)$, two basis functions, $\{t\exp(-\omega_ct), \exp(-\omega_c t)\}$,
are considered based on its analytical form. Due to the nature of a high-dimensional fitting
problem, an appropriate choice is required for the initial trial values of
$\{\omega_{\rmX;n},\gamma_{\rmX;n},\Gamma_{\rmX;n}\}$~\cite{Duan2017:PRB,Duan2017:JCP,Wang2019:JCP}.
In short, the non-monotonic variation of $C_X(t)$ in the short time
can be fitted by oscillatory or nonoscillatory terms with similar decay rates,
while the power-law decay in the long time is effectively fitted by
a series of exponential decays over different time scales.

\begin{figure}[tp]
\includegraphics[width=0.65\columnwidth]{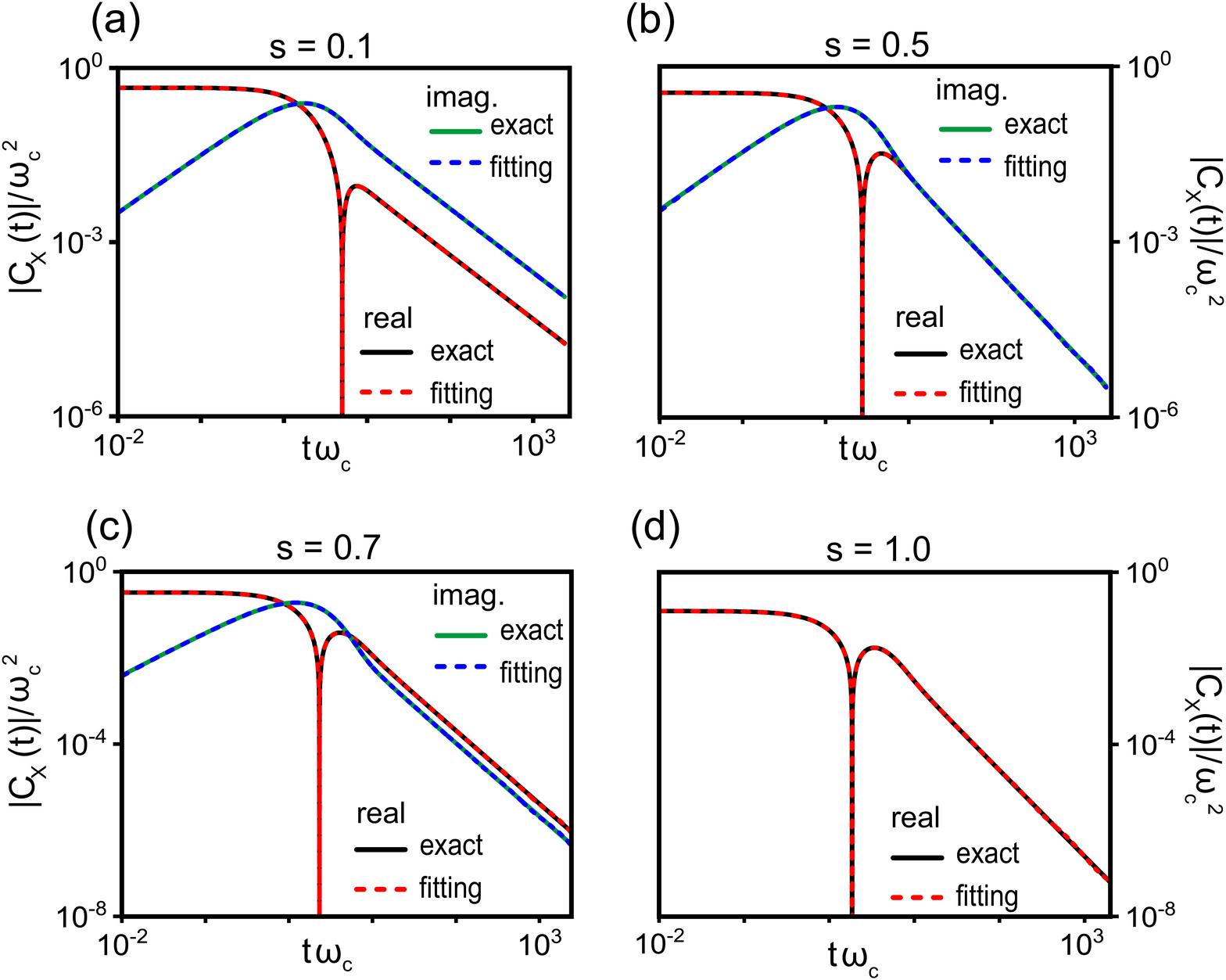}
\caption{The log-log plots of the bath correlation functions.
The bath exponents in four panels are (a) $s\!=\!0.1$, (b) $s\!=\!0.5$, (c) $s\!=\!0.7$ and (d) $s\!=\!1$.
The dashed and solid lines refer to the fitting results and the exact values, respectively.
The black and red lines denote the real parts, while the green and blue lines denote the imaginary parts.
The Kondo parameter is set to be $\alpha \!=\! 0.5$.}
\label{fig:03}
\end{figure}

As a demonstration, we present the fitting results of $C^\mathrm{fit}(t)$
for sub-Ohmic (e.g., $s\!=\!0.1,0.5,0.7$) and Ohmic ($s\!=\!1$) in Figs.~\ref{fig:02} and~\ref{fig:03}.
The Kondo parameter is set to be $\alpha\!=\!0.5$ and the maximum time is around $t_{\max}\omega_c\approx 2400$.
For $s\!=\!0.1$, the real part $C_{\rmR}(t;s\!=\!0.1)$ is fitted with 9 non-oscillatory
exponentially decaying functions. The four basis functions associated with large decay rates,
$\Gamma_\rmR/\omega_c\!=\!\{3.953, 1.059, 1.001, 0.34\}$, describe the short-time behavior ($t\omega_c\!\lesssim\!5$).
As shown in Fig.~\ref{fig:02}(a), the amplitude of the instantaneous error $\delta C_{\rmR}(t) \!=\! C_{\rmR}^{\fit}(t) - C_{\rmR}(t)$
is consistently suppressed below $|\delta C_{\rmR}(t)|/\omega^2_c\!<\!5\times10^{-5}$.
The other five basis functions associated with small decay rates,
$\Gamma_\rmR/\omega_c\!=\!\{0.112, 3.66\!\times\!10^{-2}, 1.149\!\times\!10^{-2}, 3.2\!\times\!10^{-3},5.4\!\times10^{-4} \}$,
simulate the power-law decay $C_\rmR(t) \sim -(t\omega_{c})^{-(1+s)}$ for $t\lesssim t_{\max}$.
As shown by the log-log plot of $|C_{\rmR}(t)|/\omega^2_c$ in Fig.~\ref{fig:03}(a),
this power-law decay is excellently reproduced. The imaginary part
$C_\rmI(t;s=0.1)$ is fitted with 10 basis functions. The short time ($t\omega_{c}\lesssim 5$)
behavior is fitted by two oscillatory exponentially decaying function with
$\{\omega_\rmI/\omega_c\!=\!2.108, \gamma_\rmI/\omega_c\!=\!0.793\}$
and three non-oscillatory functions with $\Gamma_\rmI/\omega_c\!=\!\{8.818, 2.102, 0.250\}$.
The long time ($t\lesssim t_{\max}$) behavior is fitted by
5 non-oscillatory functions with $\Gamma_\rmI/\omega_c\!=\!\{8.8 \times
10^{-2},3.05\!\times\!10^{-2},1.01\!\times\!10^{-2},2.97\!\times\!10^{-3}, 5.0\!\times\!10^{-4}\}$.
The excellent fitting effect is reflected by the instantaneous error $|\delta C_{\rmI}(t)|/\omega^2_c=|C^\mathrm{fit}_{\rmI}(t)-C_{\rmI}(t)|/\omega^2_c$
in Fig.~\ref{fig:02}(a) and the log-log plot of $|C_{\rmI}(t)|/\omega^2_c$ in Fig.~\ref{fig:03}(a).
For the other three bath exponents ($s=0.5$, $0.7$ and $1$), the same fitting strategy
is applied with $N_\rmR=9$ and $N_\rmI=10$ except for $C_{\rmI}(t; s=1)$.
As shown in Figs.~\ref{fig:02}(b)-\ref{fig:02}(d), the
relative instantaneous errors $|\delta C_{\rmI}(t)|/\omega^2_c$ are largely suppressed below
$5\times 10^{-5}$. As shown in Figs.~\ref{fig:03}(b)-\ref{fig:03}(d),
the long-time power-law scaling, $C_\rmX(t)\sim -(t\omega_c)^{-(1+s)}$, is also successfully duplicated.
However, we would like to emphasize that the number of basis functions needs to gradually increase
with the increase of the Kondo parameter $\alpha$ and the cutoff frequency $\omega_{c}/\Delta$ for the
final accuracy of the extended HEOM.

\subsection{Evolution of the Auxiliary Density Operators}
\label{sec3b}

Following the above decomposition of the bath correlation function, we can construct the HEOM from different approaches.
Instead of a rigorous derivation for an arbitrary initial condition~\cite{Tanimura1989:JPSJ, Tanimura1990:PRA, Tanimura2005:JPSJ, Tanimura2014:JCP, Tanimura2015:JCTC, Shao2006:CP, Shao2008:JCP, Yan2005:JCP, Yan2007:JCP, Yan2014:JCP, Moix2013:JCP, Cao2018:JCP1, Wu2015:eHEOM, Duan2017:PRB, Shi2009:JCP, Shi2014:JCP}, we present a simplified
description for a system-bath factorized initial state, $\rho(0)\!=\!\rho_\rmS(0)\otimes \rho_{\rmB}$ with
$\rho_\rmS(0)$ the initial RDM of the spin system. At a general $h(\geq 0)$-th hierarchical order, an ADO is defined as ~\cite{Wu2015:eHEOM,Duan2017:PRB,Duan2017:JCP}
\be
&&\sigma_h^{\left( \substack{
     n_1, \cdots, n_k\\
     m_1, \cdots, m_l
   } \right)} (t) = \calU_{\rmS} (t) \calT_+ \Big\{ \intt d \tau_1 \varphi_{\rmR ; n_1} (t - \tau_1) [- i\calL_z(\tau_1)] \no \\
  & &~~~\times \cdots \intt d \tau_k \varphi_{\rmR ; n_k} (t - \tau_k) [- i\calL_z (\tau_k)] \times \int_0^t d \tau_1^\pr \varphi_{\rmI ; m_1} (t - \tau_1^\pr)
      \calS_z (\tau_1^\prime) \no \\
  & &~~~\times\cdots \intt d \tau_l^\pr\varphi_{\rmI ; m_l} (t - \tau_l^\pr) \calS_z (\tau_l^\pr) \calU_{\RDM} (t) \Big\} \rho_\rmS(0).
\label{eq:12}
\ee
This ADO is characterized by the two sequences, $\{n_1, \cdots,n_k \}$ and $\{m_1, \cdots, m_l \}$,
where each index refers to a specific basis function, i.e., $n_i\rtar
\varphi_{\rmR;n_i}(\tau_i)$ and $ m_j \rtar \varphi_{\rmI;m_j}(\tau_j)$.
The hierarchical expansion order is given by the total number of basis functions, satisfying $h\!=\!k+l$.
The zeroth order ADO recovers the RDM of the spin system, giving $\sigma_0(t)\!=\!\rho_\rmS(t)$.
In Eq.~(\ref{eq:12}), $\calU_\rmS(t)\!=\!\exp{(-i\calL_\rmS t)}$ is the time propagator of the system
according to the commutator of the system Hamiltonian $\calL_\rmS\!=\![H_\rmS, \dots]$.
With respect to the Pauli-$Z$ matrix, we introduce its commutator $\calL_z(t)\!=\![\sigma_z(t), \cdots]$
and its anti-commutator $\calS_z(t)\!=\![\sigma_z(t),\cdots]_{+}$ with
$\sigma_z(t)=\calU_{\rmS}^{\dagger}(t)\sigma_{z}$.
With the consideration of the forward time ordering operator $\mathcal T_+\{\cdots\}$, we introduce an abbreviation
for a time ordering expansion, $\calU_{\RDM}(t)\!=\!\calT_+ \exp{[-\int_0^t \calW(\tau)\,d\tau]}$, which can be viewed
as a formal expression of the time propagator of the RDM. Here the transition rate kernel reads
\be
\calW(t)=\int_0^t d\tau \left[ \calL_z(t)C_\rmR(t-\tau)\calL_z(\tau) + i\calL_z(t)C_\rmI(t-\tau)\calS_z(\tau)\right].
\label{eq:13}
\ee

Next the time evolution of the $h$-th order ADO
$\sigma_h^{\left( \substack{ n_1, \cdots, n_k\\ m_1, \cdots, m_l } \right)}(t)$ is determined by the time
differentials over various $t$-dependent terms in Eq.~(\ref{eq:12}). The time derivative of each basis
function in the decomposition of $C_{\rmX=\rmR,\rmI}(t)$ is in a closed form, i.e.,
$\dot{\varphi}_{\rmX;n}(t) = \sum_{n^\pr} \eta_{\rmX;n, n^\pr}\varphi_{\rmX;n^\pr}(t)$ with
$\{\eta_{\rmX; n, n^\pr}\}$ the time-independent coefficients. Without further details, we summarize the final
time evolutions equation as
\be
&&\partial_{t}\sigma_h^{\left(\substack{n_1, \cdots,n_k\\m_1, \cdots,m_l}\right)}(t)  =  -i\calL_\rmS\sigma_h^{\left(\substack{n_1, \cdots, n_k\\m_1,\cdots, m_l}\right)}(t) \no \\
 &&~~~+\sum_{j=1}^k\sum_{j^\prime=1}^{N_\rmR}\eta_{\rmR;jj^\prime}\sigma_h^{\left(\substack{\cdots, n_{j-1}, n_{j^\prime},n_{j+1}, \cdots\\ \cdots} \right)}(t) + \sum_{j=1}^l\sum_{j^\prime=1}^{N_\rmI}\eta_{\rmI;jj^\prime}\sigma_h^{\left(\substack{\cdots\\\cdots, m_{j-1}, m_{j^\prime}, m_{j+1}, \cdots}\right)}(t) \no\\
 &&~~~-i\calL_z\sum_{j=1}^{k}\varphi_{\rmR;n_j}(0)\sigma_{h-1}^{\left(\substack{\cdots, n_{j-1}, n_{j+1}, \cdots\\\cdots} \right)}(t)
   + \calS_z\sum_{j=1}^{l}\varphi_{\rmI;m_j}(0)\sigma_{h-1}^{\left(\substack{\cdots \\ \cdots, m_{j-1}, m_{j+1}, \cdots} \right)}(t) \no\\
 &&~~~-i\calL_z\sum_{n_{k+1}=1}^{N_\rmR}a_{\rmR;n_{k+1}}\sigma_{h+1}^{\left(\substack{\cdots,n_k,n_{k+1}\\ \cdots} \right)}(t)-i\calL_z\sum_{m_{l+1}=1}^{N_\rmI}a_{\rmI;m_{l+1}}\sigma_{h+1}^{\left(\substack{\cdots\\ \cdots,m_l,m_{l+1}}  \right)}(t),
\label{eq:14}
\ee
which is an extension of the original HEOM~\cite{Wu2015:eHEOM,Duan2017:PRB,Duan2017:JCP}.
For convenience, we introduce a matrix form to rewritten Eq.~(\ref{eq:14}) into
\be
\dot{\bm{\sigma}}(t) \!=\! -\bm{\calW}\bm{\sigma}(t),
\label{eq:15}
\ee
where the vector $\bm{\sigma}(t)$ is the set of ADOs,
$\bm{\sigma}(t) \!=\! \{\sigma_{0}(t)=\rho_{\rmS}(t),\sigma_{1}(t),\sigma_{2}(t),\cdots\}$,
and the transition rate matrix $\bm{\calW}$ follows a block tri-diagonal form,
$\bm{\calW}_{h,h^{\prime}} \!=\!
\bm{\calW}_{h,h}\delta_{h^{\prime},h}+\bm{\calW}_{h,h\pm 1}\delta_{h^{\prime},h\pm 1}$.
Due to the assumptions of the system-bath factorized initial state, the initial values of the ADOs
are set to be $\bm{\sigma}(0)\!=\!\{\sigma_{0}(0)=\rho_{\rmS}(0),\sigma_{1}(0)=0,\cdots\}$.
In practice, Eq.~(\ref{eq:15}) is truncated at an expansion order $H$ for the numerical convergence.

\subsection{Application to the Linear Absorption Spectrum}
\label{sec:3c}

Although the ADOs in Eq.~(\ref{eq:12}) are obtained under the assumption of the initial system-bath factorized
state, the HEOM in Eqs.~(\ref{eq:14}) and~(\ref{eq:15}) can be applied to a much more general scenario.  The
set of the ADOs $\bm\sigma(t)=\{\sigma_0(t), \sigma_1(t), \cdots\}$ can be viewed as a complete dynamic basis
set with respect to the partial trace $\Tr_\rmB\{\cdots \rho_\rmB\}$. A one-to-one mapping can be built to
represent a reduced dynamic variable of the system in the framework of the ADOs, e.g., $\Tr_\rmB\{\cdots
e^{-i\calL t}\cdots e^{-i\calL t^\prime}\rho(0)\} \Leftrightarrow \bm{\calP}_{\rmS}\cdots e^{-\bm{\calW}
t}\cdots e^{-\bm{\calW} t^\pr}\bm{\sigma}(0)$.  Here we introduce a projection matrix onto the subspace of the
system RDM, given by
\be
\bm{\calP}_{\rmS} =
  \begin{pmatrix}
    \calI_{4} &0 &\cdots\\
    0 & 0 & \cdots\\
    \vdots & \vdots & \ddots
  \end{pmatrix}.
\ee
The block matrix in Eq.~(\ref{eq:15}) is expanded over the set of the ADOs and $\calI_4$ is
a $4\times4$ identity matrix regarding the Liouville space of the spin system.
Without further details, the linear response function in Eq.~(\ref{eq:07}) can be re-expressed in terms of the
HEOM matrices as
\be
\chi(t) &=& \lim_{t^{\prime}\rightarrow\infty}i\Uptheta (t) \Tr_{\rmS}\{\sigma_{z}\bm{\calP}_{\rmS}e^{-\bm{\calW}t}\bm{\calL}_ze^{-\bm{\calW}t^{\prime}}\bm{\sigma}(0)\},
\label{eq:16}
\ee
where the Liouville superoperator $\mathcal L_z$ is expanded in the dynamic space of the ADOs, given by
\be
\bm{\calL}_{z} =
\begin{pmatrix}
  \calL_{z} & 0 & \cdots \\
  0  & \calL_{z} & \cdots \\
  \vdots & \vdots & \ddots
\end{pmatrix}.
\ee

Following the same strategy in Sec.~\ref{sec2b}, the calculation of $\chi(t)$ is divided into four steps.
(1) With respect to a system-bath factorized initial state $\bm{\sigma}(0)$, the ADOs are propagated
by the HEOM in Eq.~(\ref{eq:15}) up to a long time ($ t^\prime \Delta\gg 1$), approaching to an approximate
equilibrium state, $\bm{\sigma}_\eq\approx\lim_{ t^\prime \Delta\gg 1}\exp(-\bm\calW t^\prime)\bm{\sigma}(0)$.
In the delocalized phase with a single ground state (equivalent to the equilibrium state at $T=0$),
$\bm{\sigma}_\eq$ is independent of the initial ADOs $\bm{\sigma}(0)$.
(2) The commutator of the Pauli-$Z$ matrix is applied onto $\bm{\sigma}_\eq$, leading to a set of
the initial quasi-ADOs, $\bm{\varsigma}(0)=\bm{\calL}_{z}\bm{\sigma}_{\eq}$.
(3) The quasi-RDM $\varrho_\rmS(t)$ is obtained by the projection of the quasi-ADOs evolved by the HEOM,
given by $\varrho_\rmS(t)=\bm{\calP}_{\rmS}e^{-\bm{\calW}t}\bm{\varsigma}(0)$.
(4) The linear response function is estimated by $\chi(t)=i\Uptheta (t)
\Tr_{\rmS}\{\sigma_{z}\varrho_\rmS(t)\}$. The linear absorption spectrum is extracted from the
imaginary part of the Fourier transform of $\chi(t)$ by Eq.~(\ref{eq:08}).

\section{NUMERICAL RESULTS}
\label{sec4}

In this section, we provide the numerical calculation of the linear absorption spectra
of the spin-boson model via the extended HEOM for the sub-Ohmic and Ohmic baths.
The cutoff frequency is set to be $\omega_{c}/\Delta\!=\!20$.
As shown in Sec.~\ref{sec3a}, the decomposition of the bath correlation function
is restricted to a maximum time $t_{\max}\omega_{c}\approx 2400$ or
equivalent to $t_{\max}\Delta \approx 120$. The numbers of the basis functions
in the decomposition are $N_\rmR\!=\!9$ and  $N_{\rmI}\!=\!10$.
The maximum expansion order of the follow-up extended HEOM is $H\!=\!12$.

\subsection{Relaxation to Equilibrium and Phase Diagram}
\label{sec4a}

In our first step of numerical calculation, we propagate
the spin-boson model from a factorized initial state 
$\rho(0)=\rho_S(0)\rho_\rmB=|+\rangle\langle+|\rho_\rmB$ to its equilibrium state $\rho_\eq$.
In the framework of the HEOM, the initial set of the ADOs is set to be
$\bm{\sigma}(0)\!=\!\{\sigma_{0}(0)\!=\!\rho_{\rmS}(0),\sigma_{1}(0)\!=\!0,\cdots\}$. 
With a long propagation time $t_\mathrm{max}\Delta\approx120$,
the evolution of $\lim_{t^\pr\rightarrow t_{\max}}\bm{\sigma}(t^\pr)=\exp(-\bm{\calW}t^\pr)\bm{\sigma}(0)$
provides an approximate but reliable estimation of $\bm{\sigma}_\eq$ in the delocalized phase.

\begin{figure}[tp]
\centering
\includegraphics[width=0.70\columnwidth]{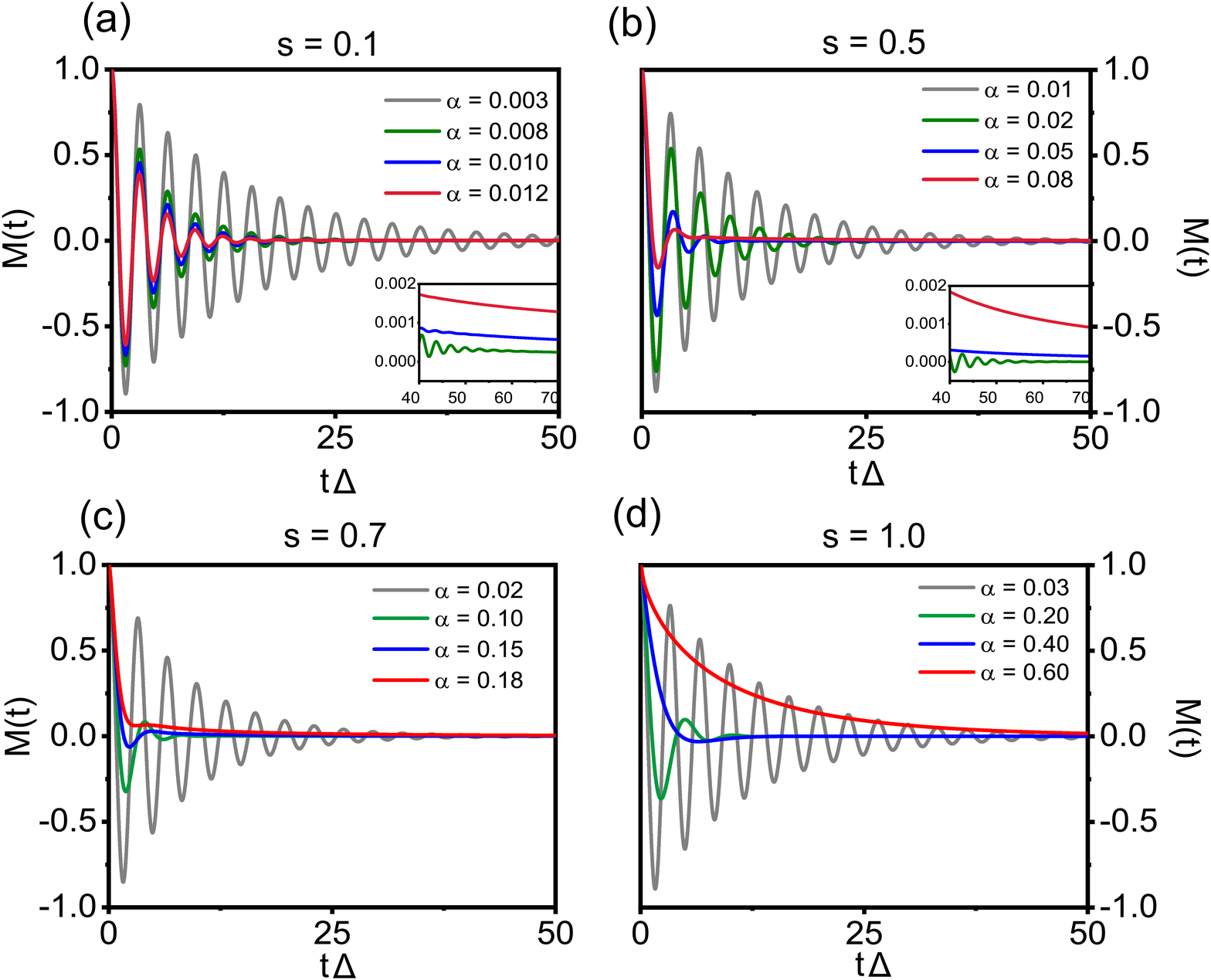}
\caption{The time evolution of the average magnetic moment in the spin system.
The bath exponents are
(a) $s \!=\! 0.1$, (b) $s \!=\! 0.5$, (c) $s \!=\! 0.7$, (d) $s \!=\! 1$.
In each panel, the four curves in colors of gray, blue, green and red
correspond to the increase of Kondo parameter.}
\label{fig:04}
\end{figure}

To visualize this time evolution, we calculate the average magnetic moment
$M(t^\pr)\!=\!\Tr_{\rmS}\{\sigma_z\bm{\calP}_{\rmS}\exp(-\bm{\calW} t^\pr)\bm{\sigma}(0)\}$
of the spin system and present the numerical results of four bath exponents,
$s=0.1$, $0.5$, $0.7$ and $1.0$, in Fig.~\ref{fig:04}.
The $\alpha$-dependence of $M(t^\pr)$ can be categorized into deep sub-Ohmic, Ohmic, and transitionary regimes.
In the deep sub-Ohmic regime ( e.g., $s \!=\! 0.1$ in Fig.~\ref{fig:04}(a) ), 
the average magnetic moment $M(t^\pr)$ shows an underdamped motion as the Kondo parameter increases. 
The oscillation period remains almost unchanged as $\pi/\Delta$.
In the short time, although $M(t^\pr)$ decays faster,
the underdamped oscillation still exist even for a strong system bath coupling, e.g., $\alpha \!=\! 0.012$.
Meanwhile, $M(t^\pr)$ exhibits a slower decay in the long time, e.g., the inset of Fig.~\ref{fig:04}(a),
which indicates the tendency of DL phase transition.
It is expected that the DL phase transition will occur when the Kondo parameter
$\alpha$ is greater than a critical Kondo parameter $\alpha_c$ and $M(t^\pr\rightarrow\infty)$ will
keep invariant on a finite value~\cite{Duan2017:PRB}.

In the Ohmic regime ( e.g., $s \!=\! 1.0$ in Fig.~\ref{fig:04}(d) ),
there is a obvious CI dynamic transition in the delocalized regime.
For the weak coupling strength, e.g., $\alpha \!=\! 0.03$, $M(t^\pr)$ also oscillate in an underdamped motion
due to the quantum coherence.
However, as the coupling strength increases, the oscillation period increases obviously and the slow decay
becomes dominant gradually.
When the coupling is strong enough, e.g., $\alpha\!=\!0.60$, the underdamped motion is depressed fully and
$M(t^{\pr})$ even shows a monotonic decay.

In the transitionary regime ( e.g., $s \!=\! 0.5,0.7$ in Fig.~\ref{fig:04}(b)-\ref{fig:04}(c) ),
the spin dynamic $M(t^\pr)$ shows a transition from deep sub-Ohmic to Ohmic regime.
For $s \!=\! 0.5$, it is an intermediate state  closer to the deep sub-Ohmic.
In the weak coupling regime, e.g., $\alpha \!=\! 0.01,0.02$,
the average magnetic moment $M(t^\pr)$ also exhibits the underdamped motion.
The short time oscillation period increases slowly 
as the Kondo parameter increases from $\alpha \!=\! 0.01$ to $\alpha \!=\! 0.02$,
Meanwhile, $M(t^\pr)$ behaves a faster decays.
For $s\!=\!0.7$, it's the intermediate state closer to the Ohmic regime.
The underdamped motion also exists in the weak coupling regime, e.g., the $\alpha \!=\! 0.02$.
As $\alpha$ increases, $M(t^\pr)$ also decays faster and oscillates in a longer period.
If $\alpha$ continues to increase, 
the underdamped motion of $M(t^\pr)$ will be depressed and the overdamped motion will emerge,
e.g., $\alpha \!=\! 0.18$.

To further quantify the DL phase transition, we can applying the method introduced in Ref.~\cite{Duan2017:PRB}
to obtain the phase diagram.
To obtain the critical Kondo parameter $\alpha_c$, we can extract the rate kernel $k(t)$ 
from the time-convoluted (TC) equation for $M(t^\pr)$
\be
\frac{d}{dt^\pr}M(t^\pr) = -2\int_{0}^{t^\pr} k(t^\pr-\tau) M(\tau) \,d\tau ,
\label{eq:18}
\ee
where the rate kernel $k(t^\pr)$ can be used to calculate the time-integrated rate,
$\kappa_{0} \!=\! \int_{0}^{\infty}k(t^\pr)\,dt^\pr$.
$\kappa_{0}$ monotonically decays as the Kondo parameter increase.
When $\kappa_{0}$ drops to zero ($\kappa_{0}\!=\!0$), the symmetry between the two spin states are broken
spontaneously and the equilibrium state is trapped according to the initial spin state.
In practice, a cutoff of the time-integrated rate  $\kappa_{0}\lesssim 0.02 \Delta$ is used to
obtain the critical Kondo parameter approximately.

Meanwhile, according to the results of relaxation to equilibrium, we can also explore the CI dynamic
transition in the short time, which is strongly affected by the bath type and the system-bath coupling
strength.
To verify the CI dynamic transition, on the one hand, the CI dynamic transition can be identified directly
from the short-time evolution of $M(t^{\pr})$.
On the other hand, the transition parameter $\alpha_{\CI}$ can also be estimated by taking the
Fourier transform~\cite{Duan2017:PRB} of linear response function in the frequency domain,
\be
\langle\delta M(\omega) \rangle = 2 \int_{0}^{\infty}\langle
\delta M(t^\pr)\rangle \cos(\omega t^\pr)\,dt^\pr,
\label{eq:19}
\ee
with $\langle \delta M(t^\pr)\rangle  \!=\! \langle M(t^\pr)\rangle -
\langle M(t^\pr\rightarrow\infty)\rangle$.
The coherent oscillation of $M(t^\pr)$ leads to a side peak in the frequency domain,
which is weakened with the increase of $\alpha$ and eventually disappears at $\alpha_{{\CI}}$.

\begin{figure}[htp]
\centering
\includegraphics[width=0.60\columnwidth]{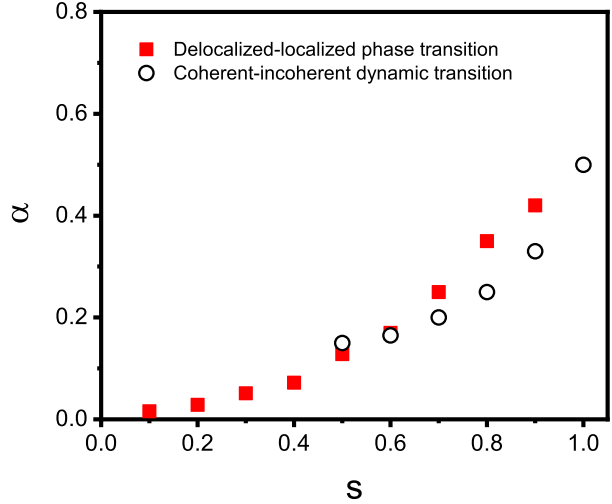}
\caption{The phase diagram of spin-boson model ($\omega_c/\Delta \!=\! 20$).
The red squares are the results of $\alpha_{c}$ from the extended HEOM.
The black circles are the results of $\alpha_{\CI}$.
}
\label{fig:05}
\end{figure}

In Fig.~\ref{fig:05}, we present the phase diagram of spin-boson model for $0.1\leqslant s \leqslant 1$.
The phase diagram can also be described by three typical regimes.
(i) Deep sub-Ohmic (e.g., $s\!=\!0.1$):
Even for the strong system-bath coupling (e.g. $\alpha\!=\!0.012$),
the coherent oscillation in short time will not disappear,
which is consistent with the spin dynamics in Fig.~\ref{fig:05}(a).
(ii)Ohmic ($s \!=\! 1$):
The transition parameter of CI dynamic transition is $\alpha_{\CI} \!=\! 0.5$.
The distance between $\alpha_{\CI}$ and $\alpha_{c}$ is very far.
Due to the computational limit, we can't obtain the exact critical parameter of DL
phase transition numerically, which is predicted as $\alpha_{c} \!=\! 1.0$ in the scaling limit.
(iii)In the transitionary state (e.g., $s \!=\! 0.5,0.7$), 
as $s$ continues to increase, the CI dynamic transition appears firstly near $s \!=\! 0.5$.
Then a crossover appears between the DL phase transition curve and the CI dynamic transition curve
near $ s\!=\! 0.6$.

\subsection{Linear Response Function and Linear Absorption Spectrum}
\label{sec4c}

Following the methodology introduced in Section.~(\ref{sec3}), next we calculate the
linear response function according to Eq.~(\ref{eq:16}) directly. The initial condition
is set as $\bm{\varsigma}(0)=\bm{\calL}_{z}\bm{\sigma}_{\eq}$, where the equilibrium state
is obtained approximately by propagating HEOM for sufficiently long period $t\Delta \!=\! 100$.

\begin{figure}[tp]
\includegraphics[width=0.7\columnwidth]{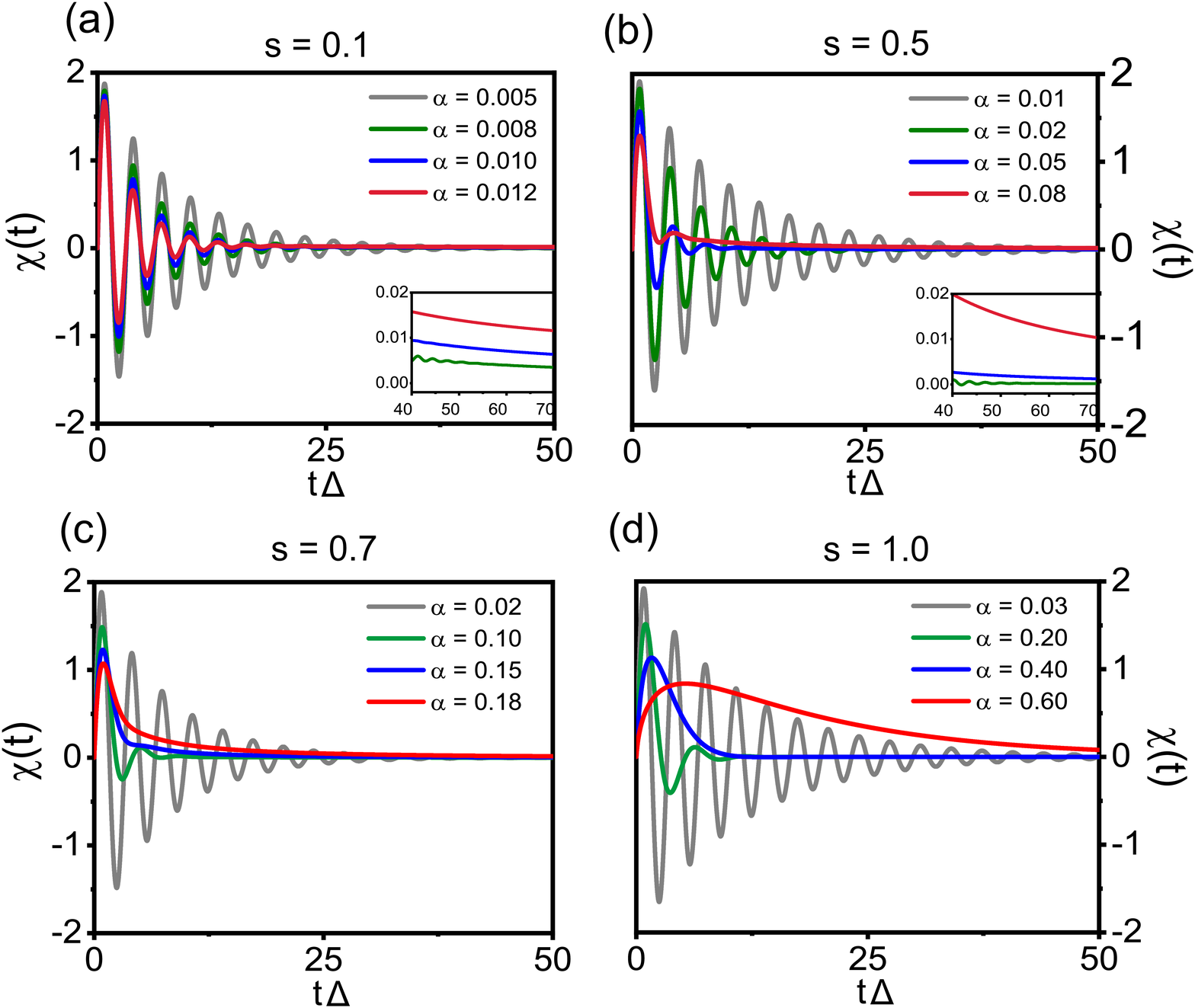}
\caption{The results of linear response function of SBM.
(a) s \!=\! 0.1,
(b) s \!=\! 0.5,
(c) s \!=\! 0.7,
(d) s \!=\! 1.
In each panel, the four curves with gray, green, blue and red,
correspond to the ascending order of Kondo parameter. All the Kondo parameters are the same
with Fig.~\ref{fig:04}.
The insets in (a)-(b) show the long-time behaviors of $\chi^{\dprime}(\omega)$.
}
\label{fig:06}
\end{figure}

In a parallel with the dynamical calculation in Fig.~\ref{fig:04}, we have
explored the linear response function over a same broad regime of bath
exponents $s$ and Kondo parameters $\alpha$.  The final result is presented in Fig.~\ref{fig:06}.
Similarly, the behavior of $\chi(t)$ can also be divided into three typical types.
(i) In Fig.~\ref{fig:06}(a), we present $\chi(t)$ at $s\!=\!0.1$ in the deep
sub-Ohmic regime with the same 4 Kondo parameters in Fig.~\ref{fig:04}(a).
In the short-time regime ($t\Delta < 20$), each $\chi(t)$ exhibits an underdamped
motion and its oscillation period is nearly unchanged with $\alpha$.
The line shape of underdamped motion can be roughly described by
$\chi(t) \approx \Uptheta (t) e^{-\gamma t} \sin(\omega t)/\omega$, where
$\gamma$ and $\omega_{0}$ are two parameters satisfying
$\omega^{2}=\omega_{0}^{2}-\gamma^{2}$ and $\gamma < \omega_{0}$.
In the long-time regime ($t\Delta > 40$), a slowly decaying tail in inset of
Fig.~\ref{fig:06}(a) is gradually developing when $\alpha$ increases from $0.008$ to $0.012$.
The line shape of overdamped motion can be roughly described by
$ \chi(t) \approx \Uptheta (t) e^{-\gamma t} \sinh(\omega t)/\omega $, where
$\gamma$ and $\omega_{0}$ are two parameters satisfying
$\omega^{2}=\gamma^{2}-\omega_{0}^{2}$ and $\gamma > \omega_{0}$.

(ii) In Fig.~\ref{fig:06}(d), we present $\chi(t)$ at $s\!=\!1$ in the Ohmic
regime. The Kondo parameters are same as Fig.~\ref{fig:04}(d).
As $\alpha$ increases from $0.03$ to $0.60$,
$\chi(t)$ exhibits a transition from underdamped motion to overdamped motion.
$\chi(t)$ even exhibits a monotonic decay soon
after a short monotonic increase (e.g., $\alpha \!=\! 0.60$ in Fig.~\ref{fig:06}(d)).
(iii) In Fig.~\ref{fig:06}(b)-(c), we present $\chi(t)$ at $s\!=\!0.5,0.7$ in the
intermediate regime between deep sub-Ohmic and Ohmic bath. 
For $s \!=\! 0.5$, $\chi(t)$ also show an underdamped motion with
same frequency for weak coupling (e.g., $\alpha \!=\! 0.01,0.02$) and a slowly monotonic decaying
(e.g., the inset in Fig.~\ref{fig:06}(b)).
From $s \!=\! 0.5$ to $s \!=\! 0.7$, the short-time underdamped motion 
is depressed gradually as the Kondo parameter increases.
For $s \!=\! 0.7$, as the coupling increases, the short-time underdamped motion even turns into a
single left-shifted peak (e.g., $\alpha \!=\! 0.18$ in Fig.~\ref{fig:06}(c)),
which is a transition signature from underdamped motion to overdamped motion.

By applying the extended HEOM kinetic method, we now have obtained the linear
response function both for Ohmic and sub-Ohmic SBM. Naturally, we 
take the Fourier transform to obtain its corresponding linear absorption
spectrum according to Eq.~(\ref{eq:08}).
At zero temperature, the linear absorption spectrum can be expanded in the sum form
as~\cite{UWeiss1985:PRL}
\be
\chi^{\dprime}(\omega)=\pi\sum_{n}|\sigma_{z_{0n}}|^{2}\delta(\omega-\omega_{n}).
\ee
where $\sigma_{z_{0n}}$ is the spectral weight of the resonance peak, which is defined as the matrix
element of $\sigma_{z}$ between $|0\rangle$ and $|n\rangle$.

\begin{figure}[tp]
\centering
\includegraphics[width=0.70\columnwidth]{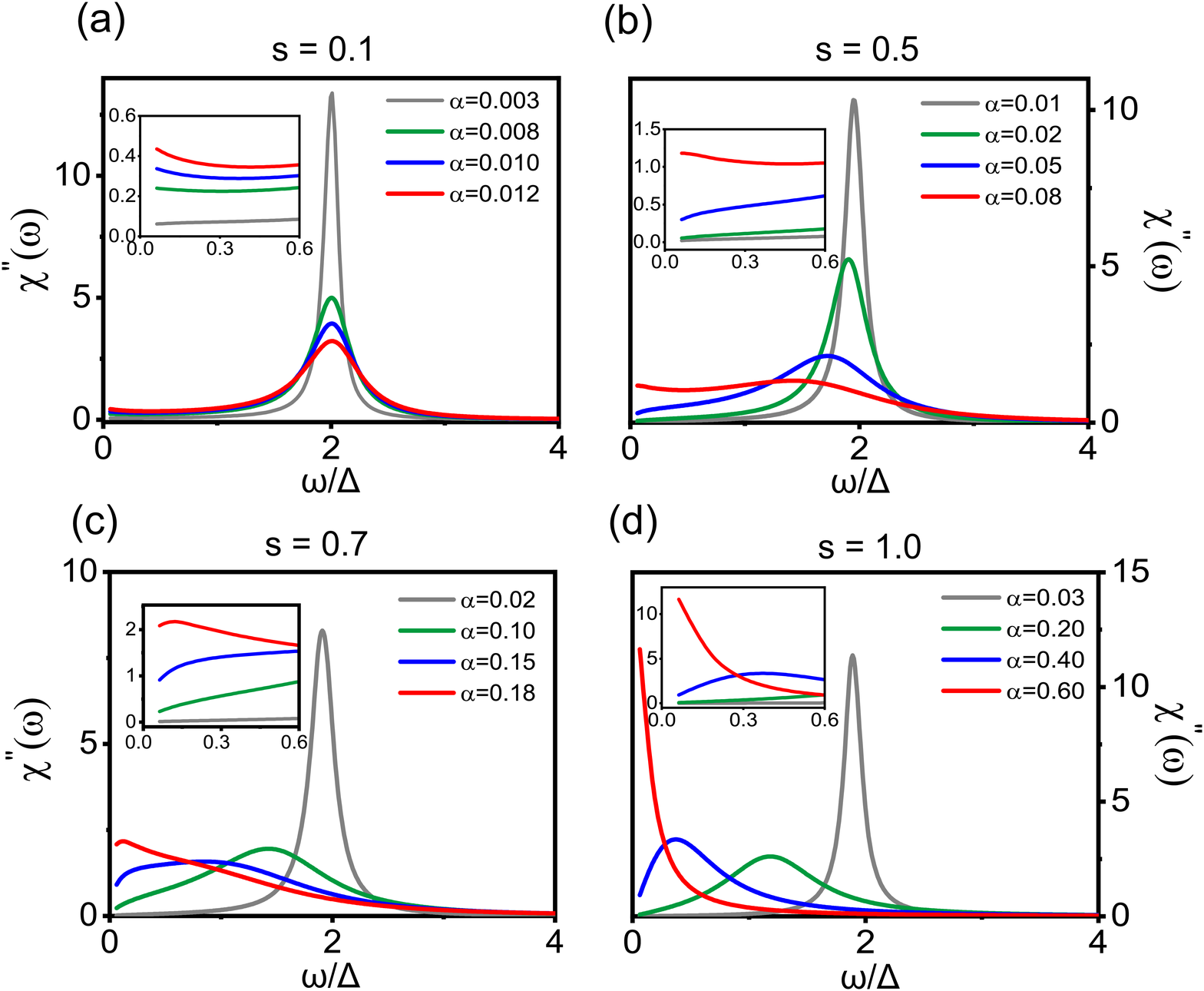}
\caption{The results of linear absorption spectrum of the sub-Ohmic and Ohmic SBM. The parameters used in the
calculation are the same with Fig.~\ref{fig:04}.
The insets show the detailed behaviors of $\chi^{\dprime}(\omega)$ near $\omega\!=\!0\Delta$.
}
\label{fig:07}
\end{figure}

As a demonstration, we also present the linear absorption spectrum $\chi^{\dprime}(\omega)$
in three typical regimes between deep sub-Ohmic and Ohmic SBM.
(i) In the deep sub-Ohmic regime (e.g., $s \!=\! 0.1$ in Fig.~\ref{fig:07}(a)),
there is always a high-frequency peak at $\omega\!=\!2\Delta$,
which corresponds to the pure underdamped motion in Fig.~\ref{fig:04}(a).
As the coupling increases, a growing low-frequency peak are found
near $\omega \!=\!0\Delta$, which corresponds to the slowly monotonic decaying in the inset of
Fig.~\ref{fig:06}(a).
(ii) In the Ohmic regime (e.g.,$s\!=\!1$ in Fig.~\ref{fig:07}(d)), their is only one absorption
peak, which reflects the renormalized tunneling between the excited state and
the ground state~\cite{Leggett1987:RMP,UWeiss1985:PRL,Zheng2009:PRE,Grifoni2018:NC}.
As $\alpha$ increases, the center of the absorption peak exhibits a tendency
shifting from $\omega \!=\! 2\Delta$ to $\omega \!=\! 0\Delta$.
Meanwhile, its height moves in a U-shape, which drops down firstly and keeps increasing subsequently.
The peak's movement is consistent with the underdamped to overdamped transition of $\chi(t)$
in Fig.~\ref{fig:06}(d).
(iii) For intermediate regime (e.g., $s \!=\! 0.5,0.7$) between the
deep sub-Ohmic and Ohmic regime, $\chi^{\dprime}(\omega)$ is presented in Fig.~\ref{fig:07}(b)-(c).
For $s \!=\! 0.5$ in Fig.~\ref{fig:07}(b),
the behavior of $\chi^{\dprime}(\omega)$ is closer to the deep sub-Ohmic.
For the weak coupling strength (e.g., $\alpha \!=\! 0.01$),
there is only one high-frequency peak near $\omega = 2\Delta$.
As $\alpha$ increases from $\alpha \!=\! 0.02$ to $\alpha \!=\! 0.08$, the center of the absorption
peak moves obviously from $\omega \!=\! 2\Delta$ to $\omega \!=\! 1.5\Delta$.
Meanwhile, the low-frequency peak with increasing height developes and shifts to zero frequency.
For $s \!=\! 0.7$ in Fig.~\ref{fig:07}(c),
in the weak coupling regime (e.g., $\alpha \!=\! 0.02$),
there is also one high-frequency peak near $\omega \!=\! 2\Delta$.
As $\alpha$ increases, it broadens and moves towards the origin.
Additionally, for a stronger coupling strength $\alpha \!=\! 0.15$,
a low-frequency peak emerges near zero.
As $\alpha$ increases from $0.10$ to $0.18$ in Fig.~\ref{fig:07}(c),
$\chi^{\dprime}(\omega)$ approaches zero frequency with increasing height.
When the coupling is strong enough (e.g., $\alpha \!=\! 0.15$),
the difference between high-frequency peak and low-frequency peak is virtually distinguishable.
The double peak structure turns into a single peak gradually
and $\chi^{\dprime}(\omega)$ behaves closer to the Ohmic case.

Additionally, all the behavior of linear absorption spectrum is consistent with the linear
response function in Fig.~\ref{fig:07}, both for sub-Ohmic and Ohmic spin-boson model.
As $\alpha$ approaches the critical Kondo parameter $\alpha \rightarrow \alpha_c$,
it is expected that a single  $\delta$-peak will appear near zero
frequency~\cite{Leggett1987:RMP, Vojta2007:PRL, Zheng2009:PRE}, $\chi^{\dprime}(\omega=0) \!=\! \infty$,
which reflects the double degenerate ground state.
The detail of $\chi^{\dprime}(\omega)$ near origin is presented in the inset of Fig.~(\ref{fig:07}).
Noticing that the minimum frequency interval in frequency domain is $0.02\pi/\Delta$.

\section{Conclusion and Disscussion}
\label{sec5}

In this paper, we apply an extended HEOM method to explore the zero-temperature spin dynamics and
linear absorption spectrum of spin-boson model from deep sub-Ohmic bath to Ohmic
bath. In order to calculate the linear absorption spectrum,
we choose to propagate SBM to its equilibrium state firstly. Once the
equilibrium state is obtained, the linear response function and its linear
absorption spectrum can be calculated subsequently according to Eq.~(\ref{eq:07}),
the Kubo formula rewritten in Liouville space.  From the time evolution of
the average magnetic moment $M(t^\pr)$, we find the spin system experiences a
dynamical transition from coherent to incoherent for a large bath exponent $s$,
which reflects the completion between the quantum coherence and decoherence induced
by the bath.

In the process of approaching to DL phase transition, we have observed the
significant difference between deep sub-Ohmic and Ohmic SBM.
Here we propose an energy level picture to explain their different behaviors.
The DL phase transition arises from the degenerate ground state.
From the view of linear absorption spectrum, in the limit $\alpha \rightarrow \alpha_{c}$,
the low-frequency absorption peak of $\chi^{\dprime}(\omega)$ approaches a $\delta(\omega)$ peak,
which is the indication of DL phase transition for Ohmic SBM.

\begin{figure}[htp]
\centering
\includegraphics[width=0.70\columnwidth]{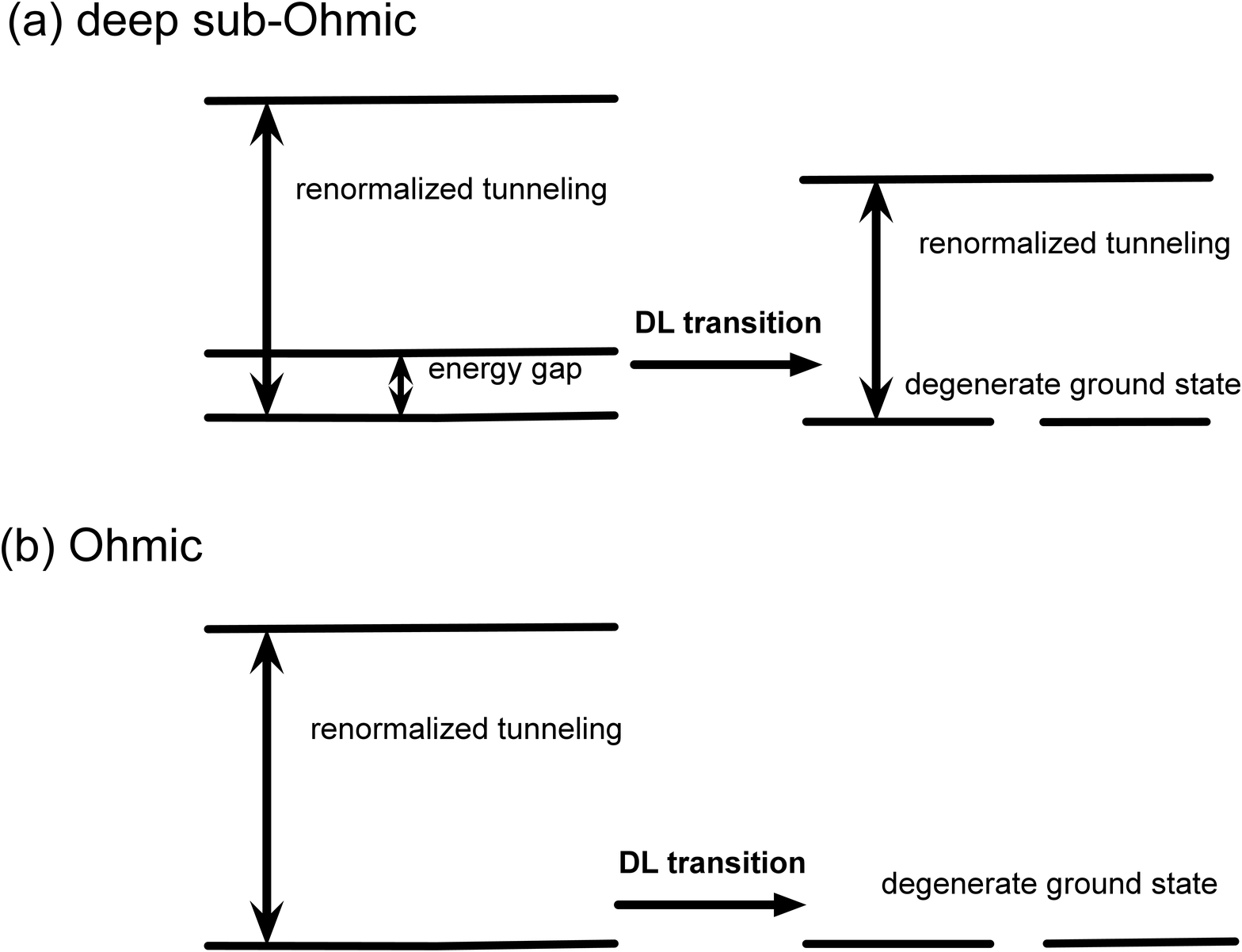}
\caption{Different pictures of the DL phase transition for deep sub-Ohmic and Ohmic SBM.
(a) deep sub-Ohmic: When the DL phase transition happpens, the two
lower levels become degenerate but their tunneling amplitudes with the highest energy level keep non-zero;
(b) Ohmic: When the DL phase transition occurs, the two energy level
becomes degenerate and the renormalized tunneling amplitude turns to zero.
}
\label{fig:08}
\end{figure}

The deep sub-Ohmic corresponds to the three-level picture in Fig.~\ref{fig:08}(a), as
it shows when the DL phase transition occurs, the overlap
integral of two lower energy levels turns to zero,
while the renormalized tunneling amplitude still remain as a finite value.
In the Ohmic regime, it corresponds to the two-level picture, so there is
only one absorption peak. When the DL phase transition
occurs, the two energy level becomes degenerate and the renormalized tunneling amplitude vanishes.

The main results include
(i) The spin dynamics $M(t)$, linear response function $\chi(t)$ and the linear
absorption spectrum $\chi^{\dprime}(\omega)$ are obtained based on extended HEOM
both for sub-Ohmic and Ohmic spin-boson model.
(ii) For the deep sub-Ohmic (e.g., $s\!=\!0.1$),
two obvious absorption peaks appear in the frequency domain. The high
frequency tunneling peak locates near $\omega\!=\!2\Delta$.
As the Kondo parameter increases, the height of high
frequency peak drops down gradually due to the renormalized effect of bath.
Meanwhile, the height of low frequency peak rises continuously, which is close to
a $\delta$-peak, when the Kondo parameter approaches the critical value, $\alpha\rightarrow\alpha_{c}$.
(iii) In the Ohmic regime ($s \!=\! 1$), there is only one robust tunneling absorption peak.
As $\alpha$ approaches the critical point $\alpha \rightarrow \alpha_{c}$,
the absorption peak with a U-shape moves to zero.
(iv) In the intermediate regime (e.g., $s \!=\! 0.5,0.7$), it's just the intermediate
state between deep sub-Ohmic and Ohmic spin-boson model.
(v) The phase diagram of delocalized-localized phase transition and
coherent-incoherent dynamic transition is obtained using the extended HEOM.
(vi) A possible explanation is proposed to understand the different mechanism
of DL phase transition between deep sub-Ohmic and Ohmic SBM.

\begin{acknowledgments}
The work reported here is supported by the Ministry of Science and Technology of
China (MOST-2014CB921203) and the National Natural Science Foundation of China (NSFC-21573195).
We would like to thank the National Supercomputer Center in Guangzhou of China for computational
support.
\end{acknowledgments}

\end{document}